\documentclass[%
 reprint,
showpacs,preprintnumbers,
nofootinbib,
 amsmath,amssymb,
 aps,twocolumn,
floatfix,
]{revtex4-1}

\usepackage{graphicx}% Include figure files
\usepackage{dcolumn}% Align table columns on decimal point
\usepackage{bm}% bold math
\usepackage{setspace}
\usepackage[latin1]{inputenc}
\usepackage{color}															 
\usepackage{colortbl}
\usepackage{enumerate}
\usepackage{framed}
\usepackage{amsmath} 
\usepackage{amsthm}
\usepackage{amssymb}
\usepackage{lineno}
\usepackage{epigraph}
\newcommand{\sign}{\text{sign}}
\begin{document}

\title{Self-organized criticality in neural networks from activity-based rewiring} 

\author{Stefan Landmann}
\author{Lorenz Baumgarten}
\author{Stefan Bornholdt} 
\email{bornholdt@itp.uni-bremen.de}

\affiliation{Institut f\"ur Theoretische Physik, Universit\"at Bremen, Bremen, Germany} 
\date{\today}
      
\begin{abstract}  
Neural systems process information in a dynamical regime between silence and chaotic dynamics. 
This has lead to the \textit{criticality hypothesis} which suggests that neural systems 
reach such a state by self-organizing towards the critical point of a dynamical phase transition. 

Here, we study a minimal neural network model that exhibits self-organized criticality 
in the presence of stochastic noise using a rewiring rule which only utilizes local information. 
For network evolution, incoming links are added to a node or deleted, depending on the node's 
average activity. Based on this rewiring-rule only, the network evolves towards a critcal state, 
showing typical power-law distributed avalanche statistics. The observed exponents are in 
accord with criticality as predicted by dynamical scaling theory, as well as with the 
observed exponents of neural avalanches. The critical state of the model is reached 
autonomously without need for parameter tuning, is independent of initial conditions, 
is robust under stochastic noise, and independent of details of the implementation 
as different variants of the model indicate. We argue that this supports the hypothesis 
that real neural systems may utilize similar mechanisms to self-organize towards 
criticality especially during early developmental stages. 
\end{abstract}


\maketitle

\section{Introduction}
Neural systems, in order to efficiently process information, have to operate at 
an intermediate level of activity, avoiding, both, a chaotic regime, as well as silence. 
It has long been speculated that neural systems may operate close to a dynamical phase 
transition that is naturally located between chaotic and ordered dynamics   
\cite{langton1990computation,herz1995earthquake,Bornholdt2000,bak2001adaptive}.  
Indeed, recent experimental results support the criticality hypothesis, most prominently 
the so-called neuronal avalanches, specific neuronal patterns in the resting state 
of cortical tissue which are power-law distributed in their sizes and durations 
\cite{Beggs2003,Beggs2004,Petermann2009,friedman2012universal,Shew2015}. 
Studies suggesting that neural systems exhibit optimal computational properties 
at criticality \cite{Larremore2011,Shew2011,Shew2009} further support the criticality hypothesis.

However, which mechanisms could drive such complex systems towards a critical state?  
Ideally, criticality is reached by a decentralized, self-organized mechanism, 
an idea known as self-organized criticality (SOC) 
\cite{bak1988self,drossel1992self,bak1993punctuated}.  
Models for self-organized criticality in neural networks were discussed even 
before experimental indications of neural criticality \cite{Beggs2003}, 
including a self-organized critical adaptive network model \cite{Bornholdt2000,Bornholdt2003}, 
as well as an adaptation of the Olami-Feder-Christensen SOC model for earthquakes \cite{olami1992}
in the context of neural networks by Eurich et al.~\cite{Eurich2002}. 

The seminal paper of Beggs and Plenz \cite{Beggs2003} eventually inspired 
a multitude of self-organized critical neural network models, often with 
a particular focus on biological details in the self-organizing mechanisms. 
Some of these mechanisms are based on short-term synaptic plasticity \cite{Levina2007}, 
spike timing dependent plasticity \cite{Meisel2009}, 
long-term plasticity \cite{Hernandez-Urbina2016}, 
while others rely on Hebbian-like rules \cite{Arcangelis2006,Pellegrini2007,Arcangelis2010} 
or anti-Hebbian rules \cite{magnasco2009self}. 
For recent reviews on criticality in neural systems see \cite{shew2013functional,markovic2014power,Hesse2015,Hernandez-Urbina2016a}.

In this paper we revisit the earliest self-organized critical adaptive network 
\cite{Bornholdt2000} in the wake of the observation of neural avalanches and ask two 
questions: Does this general model still self-organize to criticality when adapted 
to the particular properties of neural networks? How do its avalanche statistics 
compare to experimental data? Our aim is to formulate the simplest possible model, 
namely an autonomous dynamical system that generates avalanche 
statistics without external parameters and without any parameter tuning. 
\\

The original SOC network model \cite{Bornholdt2000} is formulated as a spin system, 
with binary nodes of values $\sigma \in \{-1,1\}$, corresponding to inactive and 
active states respectively. In order to study avalanches in the critical state, 
a translation to Boolean state nodes $\sigma(t) \in \{0,1\}$ is necessary, as has 
been formulated in \cite{Rybarsch2014}. That model demonstrated self-organized criticality 
based on the correlation between neighboring nodes, resulting in avalanche statistics 
in accordance with criticality \cite{Rybarsch2014}. Nevertheless, its algorithmic
implementation falls short of a fully autonomous dynamical system: Its adaptation 
rule still uses data from different simulation runs in order to determine the 
synaptic change to be performed. 
Therefore, we here formulate our model as a fully autonomous system with adaptation 
dynamics based on solely local information. It uses Boolean state nodes on a network 
without a predefined topology. The network topology changes by link adaptations 
(addition and removal of links) based on local information only, namely the 
temporally averaged activity of single nodes. Neither information of the global 
state of the system nor information about neighboring nodes, e.g.\ 
activity correlations \cite{Rybarsch2014} or retro-synaptic signals \cite{Hernandez-Urbina2016}, 
are needed. 

\section{The model}
\label{Sec:The Model}
Let us now define our model in detail. 
Consider a directed graph with $N$ nodes with binary states $\sigma(t) \in \{0,1\}$ 
representing resting and firing nodes. Signals are transmitted between nodes $i$ and $j$ 
via activating or inhibiting links $c_{ij}\in\{-1,1\}$. If there is no connection between 
$i$ and $j$ we set $c_{ij}=0$. Besides the fast dynamical variables $\sigma(t)$ of the 
network, the connections $c_{ij}$ form a second set of dynamical variables of the system 
which are evolving on a considerably slower time scale than the node states $\sigma(t)$. 
Let us define these two dynamical processes, activity dynamics and network evolution, separately. 

\subsection{Activity dynamics}
The state  $\sigma_i(t+\Delta t)$ of node $i$ depends on the input
\begin{equation}
f_i(t)=\sum\limits_{j=1}^{N}c_{ij} \, \sigma_j(t)
\end{equation}
at some earlier time $t$. For simplicity of simulation we here chose a time step of $\Delta t = 1$
and perform parallel update such that this time step corresponds to one sweep where each node is updated 
exactly once. Please note that random sequential update as well as an autonomous update of each node 
according to a given internal time scale is possible as well and does not change our results.  
Having received the input $f_i(t)$, node $i$ will be active at $t+1$ with a probability
\begin{equation}
\text{Prob}[\sigma_i(t+1)=1]=\frac{1}{1+\text{exp}[-2\beta(f_i(t)-0.5)]} \label{Eq:On}.
\end{equation}
Here, $\beta$ is an inverse temperature, solely serving the purpose of quantifying the 
amount of noise in the model. 
For the low-temperature limit $\beta \rightarrow \infty$ the probability (\ref{Eq:On}) 
becomes a step function which equals $0$ for $f_i<0.5$ and $1$ for $f_i>0.5$. 
This function broadens for decreasing $\beta$, also allowing for nodes being active once in a while 
without receiving any input. Such idling activity is observed in cortical tissue and will play a 
role in the evolutionary dynamics as defined in the following. 
\\
This model attempts to formulate the simplest rules for the activity dynamics possible, 
i.e. with the fewest states of the nodes and the fewest parameters. 
Thus the dynamics neither considers a refractory time nor a non-zero activation threshold. 
Nevertheless, as shown in Sec.~\ref{Sec:Other Versions}, the mechanism driving the network 
towards criticality works in very different biological implementations of the model. 
This suggests that despite being a coarse simplification of a real biological system, 
the model is able to represent basic mechanisms which can also be at work in the more complex real neuronal systems.

\subsection{Network evolution}
Following the natural time scale separation between fast neuron dynamics and slow change of 
their connectivity, we here implement changes of the network structure itself on a 
well-separated slow timescale. For every time step, each node is chosen with a small probability 
$\frac{\mu}{N}\ll 1$ and its connectivity is changed on the basis of its average activity 
$A_i=\left<\sigma_i \right>_W$ over the time window of the last $W$ time steps according to the following rules: 
\begin{itemize} 
		\item $A_i=0$: add a new incoming link $c_{ij}=1$ from another randomly chosen node $j$.  
		\item   $A_i=1$: add a new incoming link $c_{ij}=-1$ from another randomly chosen node $j$.       
		\item   $A_i \not\in \{0,1\}$: remove one incoming link of $i$.  
\end{itemize}
Thus, inactive (i.e.\ non-switching) nodes receive new links, while active (i.e.\ switching) nodes lose links. 
These rules prevent the system from reaching, both, an ordered phase where all nodes are 
permanently frozen, as well as a chaotic regime with abundant switching activity. 
In particular, the system is driven towards a dynamical phase transition between a globally 
ordered and a globally chaotic phase. Note that the sign of an added link is determined by 
the nature of the frozen state (on or off), unlike the original SOC network model where the 
sign of new links had been chosen randomly \cite{Bornholdt2000}. 

Note that rewiring is based on locally available information only. To simulate the way
a single cell could keep a running average, we also implemented the average activity of 
a node as $A_i(t+1)=\sigma(t+1)\cdot (1-\alpha) + A_i(t)\cdot \alpha$
as the basic principle a biochemical average would be taken. Here, the parameter 
$\alpha \in [0,1]$ determines the temporal memory of the nodes (instead of the averaging 
time window parameter $W$). Since the newly defined $A_i$ can only approach but never attain 0 or 1, 
we have to reformulate the criteria which determine the type of rewiring to be performed. 
The condition for a node to receive an activating link is transformed from $A_i = 0$ to 
$A_i < \epsilon$ with $\epsilon \ll 1$, the other criteria are changed correspondingly. 
Then, we find that the model works accordingly.

For practical purposes, we perform the rewiring of only one randomly chosen node $i$ after 
every $T=\frac{N}{\mu} $ sweeps, instead of selecting every node with a certain probability 
$\mu \over N$ at each time step. Both implementations yield the same results.

The proposed rules for the network evolution are inspired by synaptic wiring and rewiring 
as observed in early developmental stages of neural populations or during the rewiring 
of dissociated cortical cultures \cite{yada2017development}. 
In these systems, homeostatic plasticity mechanisms are at work, which lead to an increasing 
activity of overly inactive neurons and vice versa. In \cite{mattson1989excitatory} 
it was found that the application of inhibitory neurotransmitters to pyramidal neurons 
in isolated cell cultures, and thus a decrease of activity, leads to an increased outgrowth 
of neurites. In contrast, if excitatory neurotransmitters are applied, a degeneration of 
dendritic structures is induced \cite{mattson1989excitatory,mattson1988outgrowth,haydon1987regulation}.  
These observations were confirmed in experiments where electrical stimulation of neurons 
showed to inhibit dendritic outgrowth \cite{cohan1986suppression} and blocking of activity 
resulted in an increased growth of dendrites \cite{fields1990effects,van1987tetrodotoxin}. 
Thus, if a neuron is overly inactive or active, it "grows and retracts its dendrites to 
find an optimal level of input ...''  \cite{fauth2016opposing} which is mimicked by the 
proposed rewiring rules. Similar homeostatic adaption rules have been successfully used 
to model cortical rewiring after deafferentation \cite{butz2009model}.
\\

The rewiring rules of the model use information on the level of single nodes only and 
are robust, despite their simplicity. To show that the mechanism also works in different 
biological implementations, we examined modified versions of this model, 
see Sec.~\ref{SEC: Other versions} for details. For example, criticality 
is also reached if the strength of the links is allowed to assume continuous values. 
Furthermore, introducing inhibiting neurons instead of inhibiting links does not 
change the dynamics of the model. 
 
\section{Evolution of the network structure} 
\label{Sec:Structure} 
The evolution of the network starts with a specified initial configuration of links 
$\mathbf{c}(t=0)$ and the state of all nodes set to $\bm{\sigma}(t=0)=\mathbf{0}$. 
Doing so, all activity originates from small perturbations caused by stochastic noise. 
Applying the rewiring rules, the system then evolves towards a dynamical steady state 
with characteristic average numbers of activating and inhibiting links. 
 
As a convenient observable of the dynamical state of the network, and an approximate indicator of 
a possible critical state of the network, we measure the branching parameter $\left< \lambda \right>$ 
by calculating, for every node $i$, how many neighbor nodes $\lambda$ on average change their state 
at time $t+1$ if the state of $i$ is changed at time $t$. 
Averaging $\lambda$ over the network indicates the dynamical regime of the network. 
where $\left< \lambda \right> =1$ is often used as an indicator of criticality. 
Note that, by construction, $\left< \lambda \right>$ depends on the connectivity matrix 
$c_{ij}(t)$ and on the state vector $\bm{\sigma}(t)$ and, therefore, has to be considered 
with some caution. For example, its critical value may differ from one when the evolved networks 
develop community structure or degree correlations between in- and out-links or between nodes \cite{Larremore2011}. 
Therefore, we will here use the branching parameter for a qualitative assessment of the network evolution, only, 
and analyze criticality with tools from dynamical scaling theory below. 

\begin{figure*}[]
	\includegraphics[width=\linewidth]{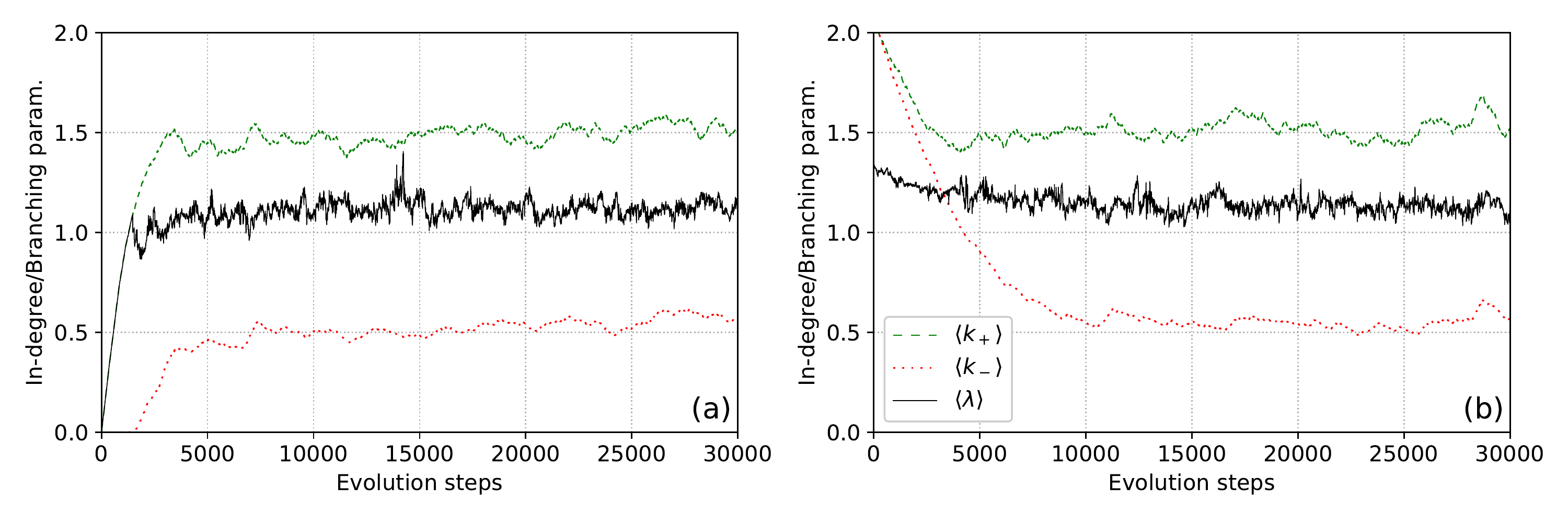}
	\caption{(Color online) (a) Time series of the average in-connectivities and branching parameter for 
	$N=1000, \, \beta=10, \,T=1000$, starting from a completely unconnected network. 
	After a transient period, the average connectivities and the branching parameter become stationary. 
	The branching parameter fluctuates around $\left< \lambda \right> =1.10 \pm 0.11$, indicating possible criticality. 
	(b)  Evolution starts with an average connectivity of $\left<k\right>=2$ for activating and inhibiting links. 
	Even though having a very different initial configuration, the system evolves towards a similar steady state 
	as found in (a).}
	\label{FIG: Evolution of links c(0)=0}
\end{figure*}

Let us now turn to the evolutionary dynamics of the model, starting from a random network $\mathbf{c}(t=0)$ 
with only the average connectivity specified at $t=0$. 
Figure~\ref{FIG: Evolution of links c(0)=0}a shows the time series of the average number of incoming 
activating and inhibiting links per node $\left< k_+ \right>$ and  $\left< k_- \right>$ starting from 
a fully unconnected network.  
The figure also shows the temporal evolution of the branching parameter $\left< \lambda \right>$. 
\\
In the beginning of the network evolution, there are only few links between the nodes, and noise induced 
activity dies out very fast. Therefore, the activity is very low and only activating links are added. 
As a result, the branching parameter increases. When the value of $\left< k_+ \right>$  approaches one, 
activity starts to propagate through the network and some nodes become permanently active. 
This causes the rewiring algorithm to insert inhibiting links. After some transient time, 
the average connectivities become stationary and fluctuate around a mean value. 
The branching parameter also becomes stationary and fluctuates around a value near one, 
indicating a possible critical behavior. The ratio of inhibiting links to activating 
links approximately attains $\left<k_-\right>{/}\left<k_+\right> \approx 0.3$ which is close to
the ratio of inhibition/activation typically observed in real neural systems \cite{Markram2004}.
The connectivity in the stationary states exhibits Poisson-distributed degree distributions of incoming and outgoing links.
\\

Figure~\ref{FIG: Evolution of links c(0)=0}b shows the evolution of the average connectivities 
with different initial conditions. Here, the initial average connectivities are chosen as 
$\left<k_+\right>=\left<k_-\right>=2$. In contrast to the starting configuration in 
Fig.~\ref{FIG: Evolution of links c(0)=0}a, the network is densely connected and the nodes 
change their states often. 
Since the nodes rarely stay in the same state during the averaging time $W$, 
links are preferentially deleted in the beginning. After a transient period the system reaches 
a stationary steady state similar to the one already observed in 
Figure~\ref{FIG: Evolution of links c(0)=0}a, indicating independence from initial conditions.
 
This scenario is reminiscent of synaptic pruning during adolescence, where in some regions of 
the brain approximately 50\% of synaptic connections are lost \cite{Rakic1994}.
It is hypothesized that this process contributes to the observed increase in efficiency of 
the brain during adolescence \cite{Spear2013}. In the proposed model, starting with the densely 
connected network shown in Fig.~\ref{FIG: Evolution of links c(0)=0}b, the branching parameter 
is considerably larger than one. In this state information transmission and processing are difficult
since already small perturbations percolate through the entire network. 
The decrease in the number of links leads to a network with a branching parameter close to one,
much better suited for information processing tasks. 

\begin{figure*}[] 
\includegraphics[width=\linewidth]{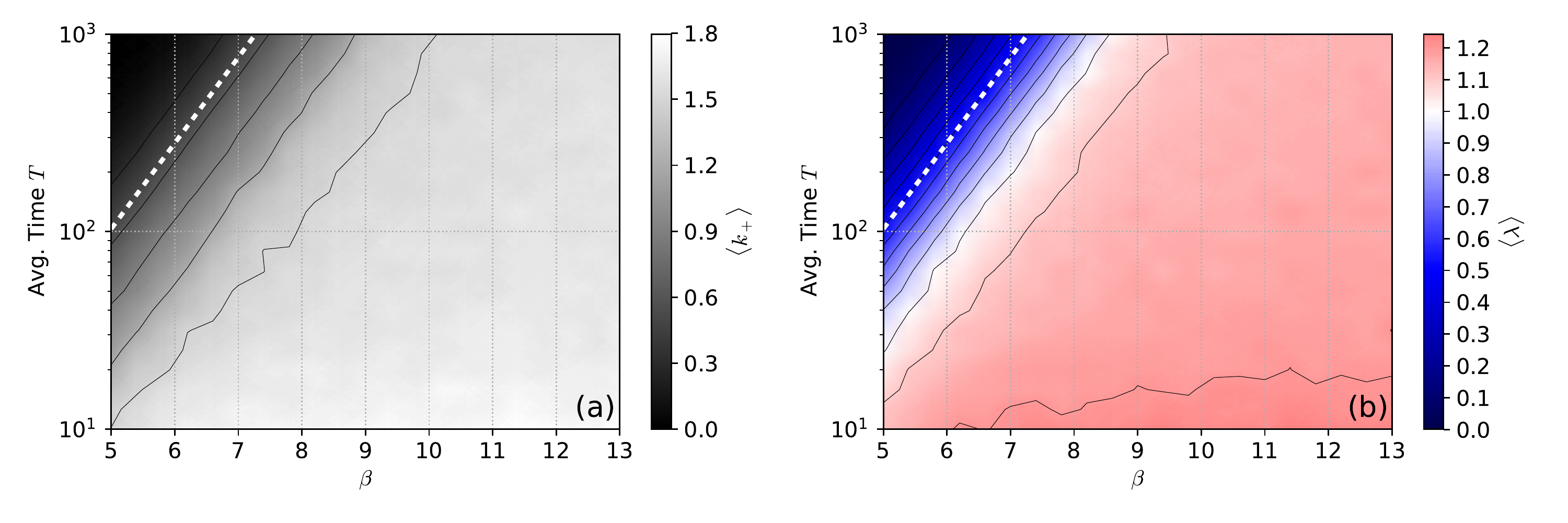} 
\caption{(Color online) (a) The average connectivity of activating incoming links $\left< k_+ \right>$ 
    for different values of the averaging length $T$ and the inverse temperature $\beta$. 
    The white dotted line is the upper bound of $T$ given by Eq.~(\ref{EQ: Upper bound W}). 
    (b) The average branching parameter $\left< \lambda \right>$ for different values of 
    the averaging window length $W$ and the inverse temperature $\beta$. 
    The average branching parameter is close to a typical value near one over a broad range of $(\beta,W)$. 
    The data was obtained by averaging over 30000 evolution steps, system size is $N=1000$.}
\label{FIG:Connectivity scan}
\end{figure*}

In order to explore the parameter dependency of the model, let us now ask how the steady-state-averages 
of the connectivities and of the branching parameter depend on the system parameters $(\beta,W,N)$. 
Figure~\ref{FIG:Connectivity scan}a shows the average connectivity of activating incoming links 
over a broad range of parameter space. A prominent feature is the sub-critical 
region (upper left corner) where the algorithm fails to create connected graphs and the average 
connectivity of incoming links is far below one. This is due to nodes being predominantly active by noise, 
instead of signal transmission. If a node $i$ has no incoming links its probability to be turned on 
at least once by noise during the $W$ time steps is given by
\begin{equation}
\text{Prob}(A_i>0)=1-\left(1-\frac{1}{1+e^\beta}\right)^W.
\end{equation}
Therefore, demanding that on average not more than half of the nodes should be turned on by noise during 
$W$ steps, gives an upper bound for the time window $W$
\begin{equation}
W_{\text{max}}=-\frac{\text{log} \, 2}{\text{log}\left(1-\frac{1}{1+e^\beta}\right)}.
\label{EQ: Upper bound W}
\end{equation}
This boundary is shown as a white dashed line in Fig.~\ref{FIG:Connectivity scan}a, 
obviously being a good approximation for the boundary of the sub-critical region. 
Most importantly, we see that if $\beta$ is sufficiently large, i.e.\ if the noise is 
sufficiently small, there always is a region in which connected networks emerge. 
Since $W_{\text{max}}$ is independent of system size $N$, this also holds for large systems.
Fig.~\ref{FIG:Connectivity scan}b shows the average branching parameter for the same range of $(\beta,W)$ 
as Fig.~\ref{FIG:Connectivity scan}a. 
Note that $\left< \lambda \right>$ is close to a value slightly larger than one, over a wide range of noise and averaging times.
In order to explore whether this indicates criticality (with a critical branching parameter value 
larger than one for the evolved networks), let us now explore other criteria of criticality. 

\section{Criticality}
\label{Sec:Indicators}
An important feature of critical systems is scale-independent behavior, 
meaning that close to a phase transition similar patterns can be observed on all scales. 
Near criticality, correlations between distant parts of the system do not vanish and 
microscopic perturbations can cause influences on all scales. 
This also implies that power-laws occur in many observables, as e.g.\ in the size 
distribution of fluctuations.
\subsection{Avalanches of perturbation spreading}
Let us now investigate the statistics of avalanches of perturbations spreading on the networks. 
Note that the network evolution drives the system towards a state where activity never dies out.  
Therefore, we cannot consider avalanches of activity-spreading, as usually done in numerical 
experiments, with one perturbation at a time.  
The problem of persistent activity could be circumvented by introducing an activity threshold 
which defines the start and the end of avalanches as done in \cite{poil2012critical}. 
This procedure, nevertheless, is not reliable since the introduction of an activity threshold 
can generate power-law-like scaling from uncorrelated stochastic processes as was shown 
in \cite{touboul2010can}. Instead, showing that the size and duration of the fluctuations 
are power-law distributed is a more reliable procedure commonly used in statistical physics 
\cite{stanley1971phase}. This method is related to the determination of the Boolean Lyapunov 
exponent, which was used e.g.\ in \cite{haldeman2005critical} in order to examine the 
critical behavior of neural networks. 
\\

First, we let the system evolve until the branching parameter and the average connectivities 
reach steady average values. Then, a copy $\bm{\sigma}_c$ of the network is made. 
One node of this copy is chosen at random and its state is flipped: if it was active, 
it is turned inactive and vice versa. By comparing the temporal evolution of the 
unperturbed system $\bm{\sigma}$ and the perturbed system $\bm{\sigma}_c$ one can 
examine the spreading of this perturbation. For quantifying the 'difference' between 
the two copies it is convenient to use the Hamming-distance of the state vectors 
$d_H(\bm{\sigma},\bm{\sigma}_c)$ which is defined as the number of differing entries 
in $\bm{\sigma}$ and $\bm{\sigma}_c$, i.e. the number of nodes which deviate from 
each other in their states. During the examination of one perturbation the rewiring 
algorithm is not in action. 
 
Performing simulations we found that in most cases $d_H(\bm{\sigma},\bm{\sigma}_c) \rightarrow 0$ 
after some time, which means that the perturbed system falls back onto the attractor 
of the unperturbed system. For a system of e.g.\ 2000 nodes with $\beta=10$ and $W=1000$ this 
was observed in more than 90~\% of all perturbations.
 
It is straightforward to define the avalanche duration $T$ as the time between the start 
of the perturbation and the return of $\bm{\sigma}_c$ to the same attractor as $\bm{\sigma}$ 
and the avalanche size $S$ as the cumulative sum of the Hamming-distances between 
$\bm{\sigma}$ and $\bm{\sigma}_c$ during the avalanche: 
\begin{equation}
 S=\sum_{t=0}^{T}d_H(\bm{\sigma}(t),\bm{\sigma}_c(t)).
\end{equation}
From universal scaling theory \cite{Sethna2001} it is expected that these observables exhibit 
power-law scaling at criticality
	\begin{align}
	\text{Prob}(S) & \thicksim S^{-\tau},\\
	\text{Prob}(T) & \thicksim T^{-\alpha}.
	\end{align}
Furthermore, it should also hold that the relation between the average avalanche size and 
the avalanche duration shows power-law scaling 
\begin{align}
	\left< S\right> (T)  \thicksim T^{-\gamma},
\end{align}
	with the exponents fulfilling the relation
	\begin{equation}
	\label{Eq:Exponen realtion}
	\frac{\alpha-1}{\tau-1}=\gamma. 
	\end{equation}
To further verify criticality it is possible to explicitly show the scale-freeness of the 
avalanche dynamics. This can be done by determining the average avalanche profiles 
(avalanche size over time) for different avalanche durations. 
Scaled properly, these shapes should collapse onto one universal curve if the system is critical.
\subsection{Results}
\begin{figure*}[]
	\begin{minipage}[t]{.95\columnwidth} 
		\includegraphics[width=0.95\columnwidth]{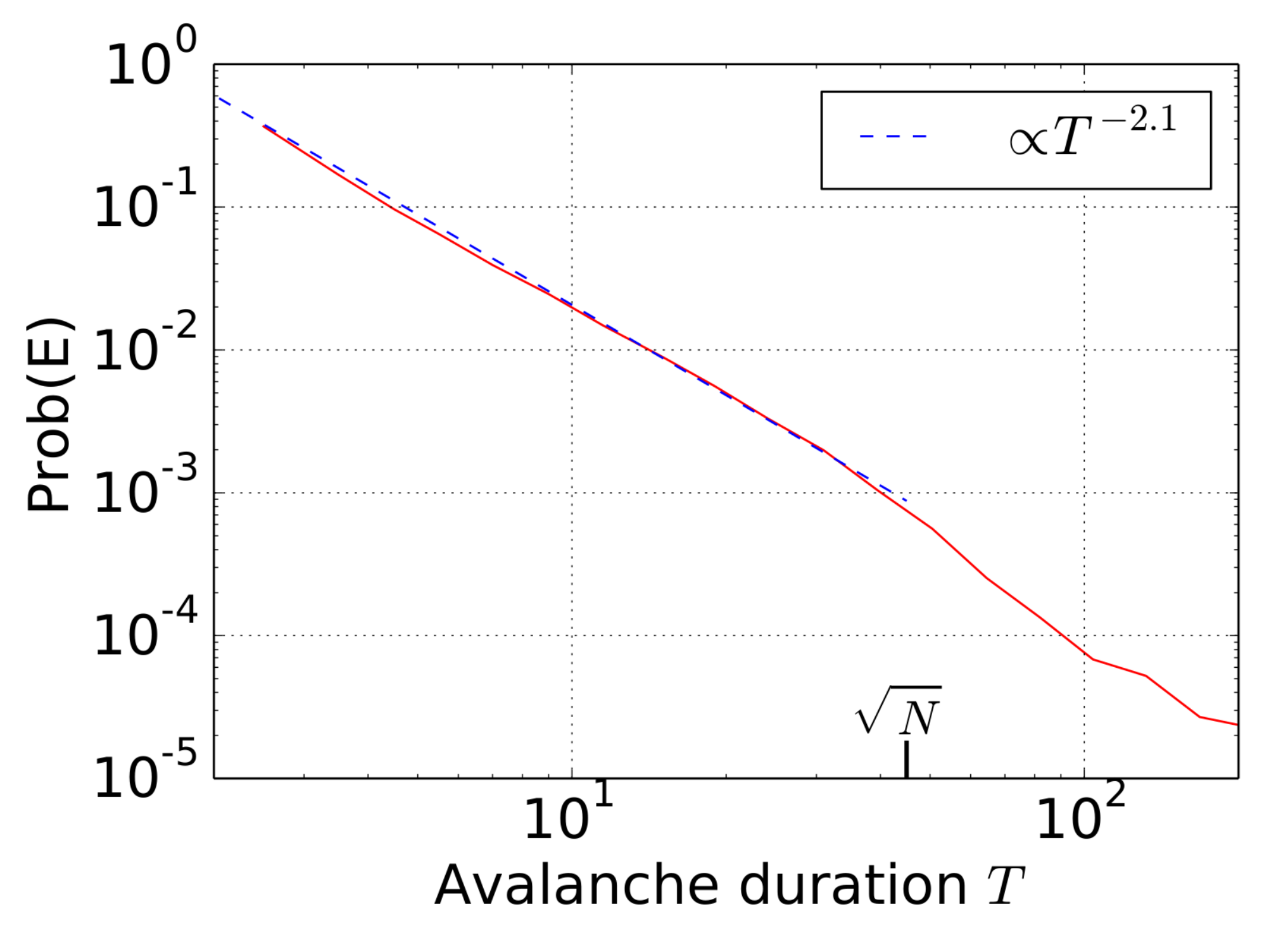}
	\end{minipage}
		\hspace{0.2cm}
	\begin{minipage}[t]{0.95\columnwidth}
		\includegraphics[width=0.94\columnwidth]{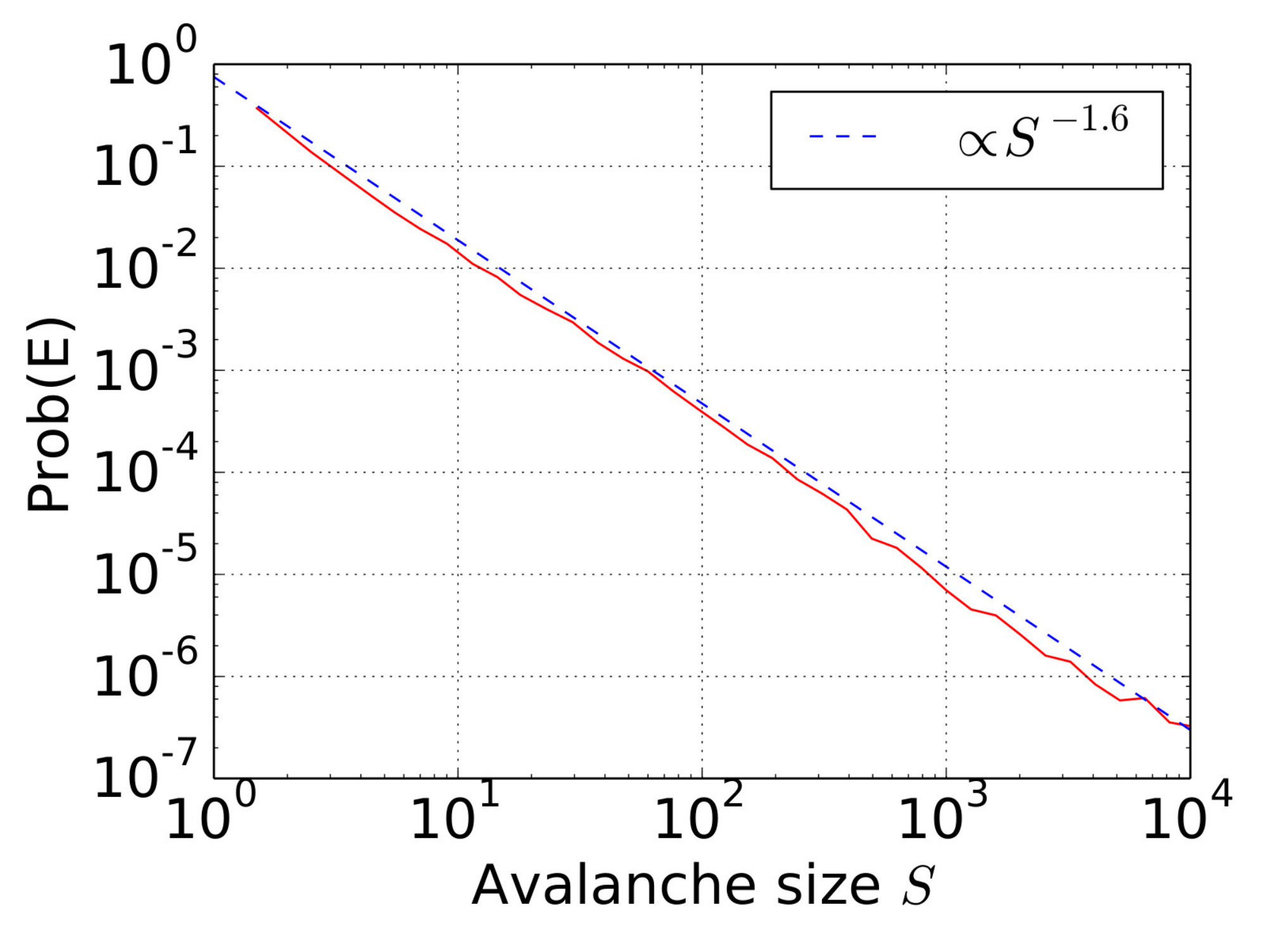}
	\end{minipage}
\hphantom{TTTTT}(a) \hspace{8cm}(b)
	\hspace{0.3cm}
	\begin{minipage}[t]{0.95\columnwidth}
	\includegraphics[width=0.94\columnwidth]{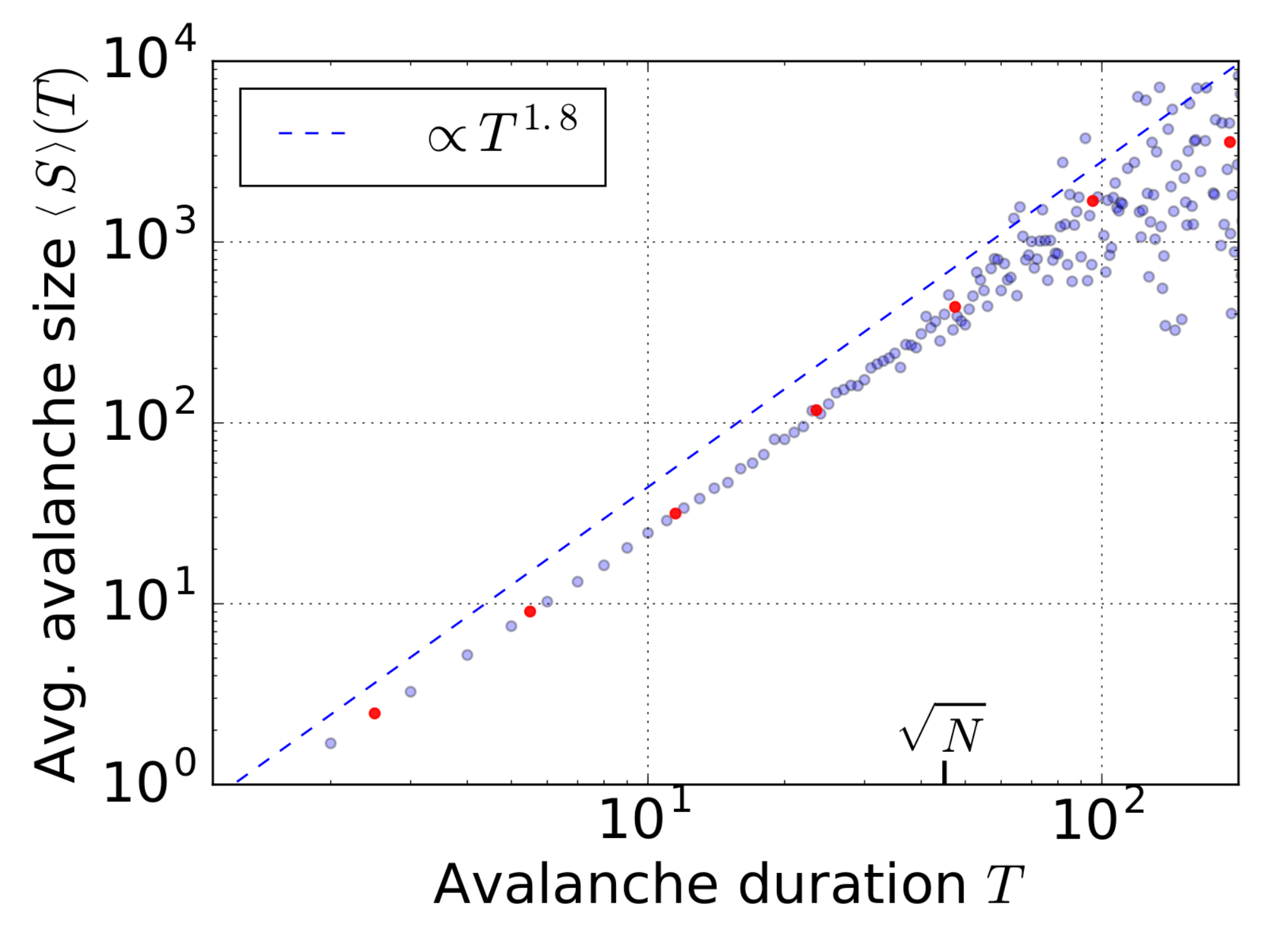}
	\end{minipage}
	\hspace{0.2cm}
	\begin{minipage}[t]{0.95\columnwidth}
	\includegraphics[width=0.93\columnwidth]{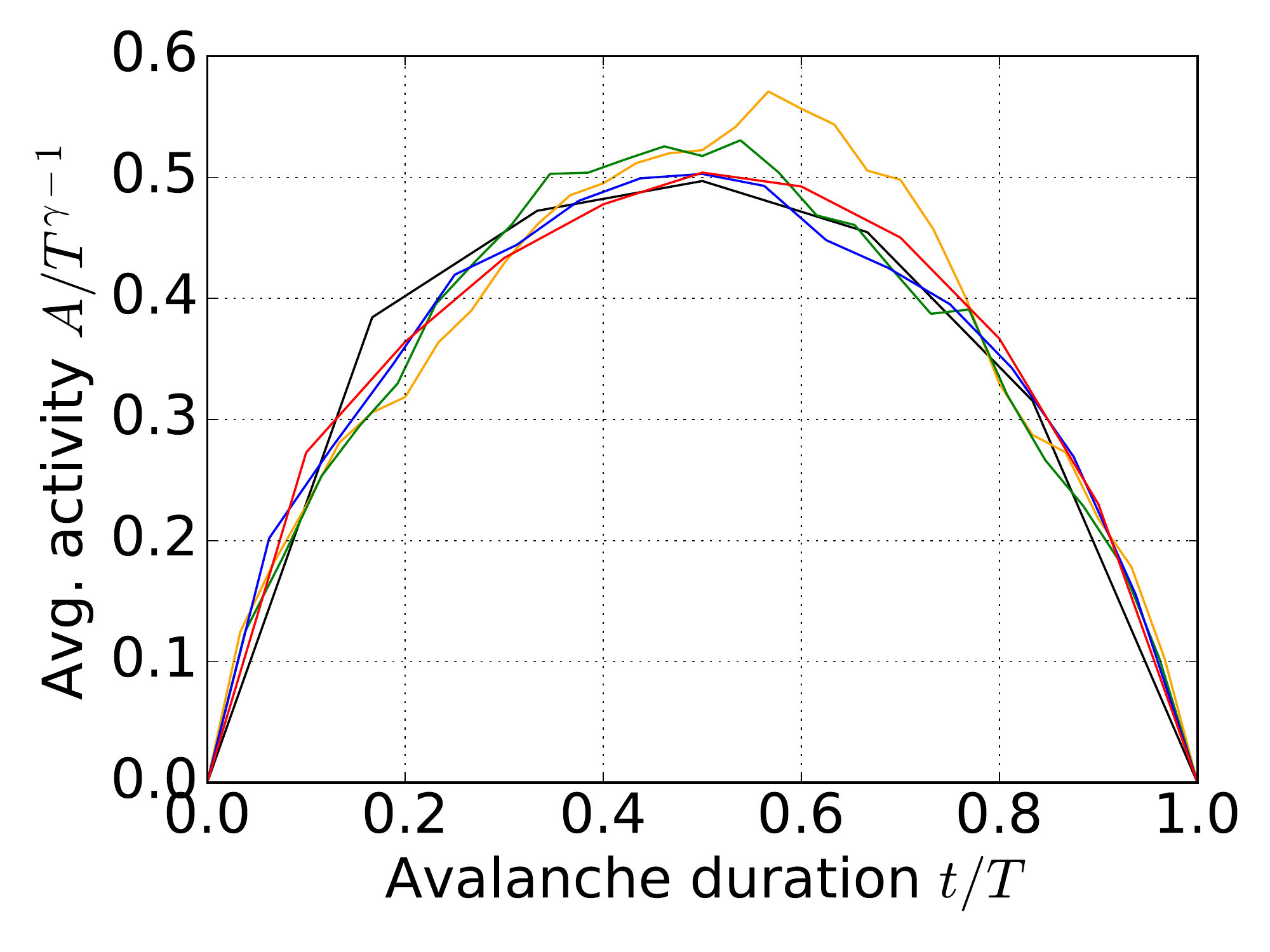}
	\end{minipage}
\hphantom{TTTTT}(c) \hspace{8cm}(d)
	\caption{(Color online) Avalanche statistics and collapse of avalanche profiles. 
	(a) Avalanche duration distribution. 
	(b) Avalanche size distribution. 
	(c) Average avalanche size over avalanche duration. The red dots show logarithmically binned data. 
	(d) Collapse of avalanche shapes. The curves show the activity during avalanches of 
	length 6 (red), 10 (black), 16 (blue), 30 (green), 36 (yellow). Parameters: 
	$N=2000, W=1000, \beta=10$, data from $75 \cdot 10^3$ avalanches. 
	}
	\label{Fig: Landmarks of criticality}
\end{figure*}

Figure~\ref{Fig: Landmarks of criticality} shows the distribution of avalanche sizes and 
durations as well as the collapse of avalanche profiles for avalanches of different durations.
\\
(a) shows that the avalanche duration scales with an exponent of $\alpha=2.05 \pm 0.03$ 
up to the square root of the system size. (b) reveals a power-law scaling of the avalanche 
size with an exponent of $\tau=1.61\pm 0.01$. Both exponents $\alpha$ and  $\tau$ are in 
line with experimental results \cite{Beggs2003,friedman2012universal}. (c) shows that 
the relation between average avalanche size and avalanche duration also exhibits 
a power-law scaling up to the square root of system size with an exponent of 
$\gamma=1.76 \pm 0.03$. These exponents fulfill the relation    
\begin{equation}
\frac{\alpha-1}{\tau-1}=\frac{1.05\pm 0.03}{0.61\pm 0.01}= 1.72 \pm 0.07 \approx \gamma
\end{equation}
strongly suggesting that the system is critical. 
(d) shows the collapse of the activity curves onto one universal shape, 
as it was also found in experiments \cite{friedman2012universal}, 
reflecting the fractal structure of the avalanche dynamics.
\\

\begin{figure}[h]
	\includegraphics[width=\columnwidth]{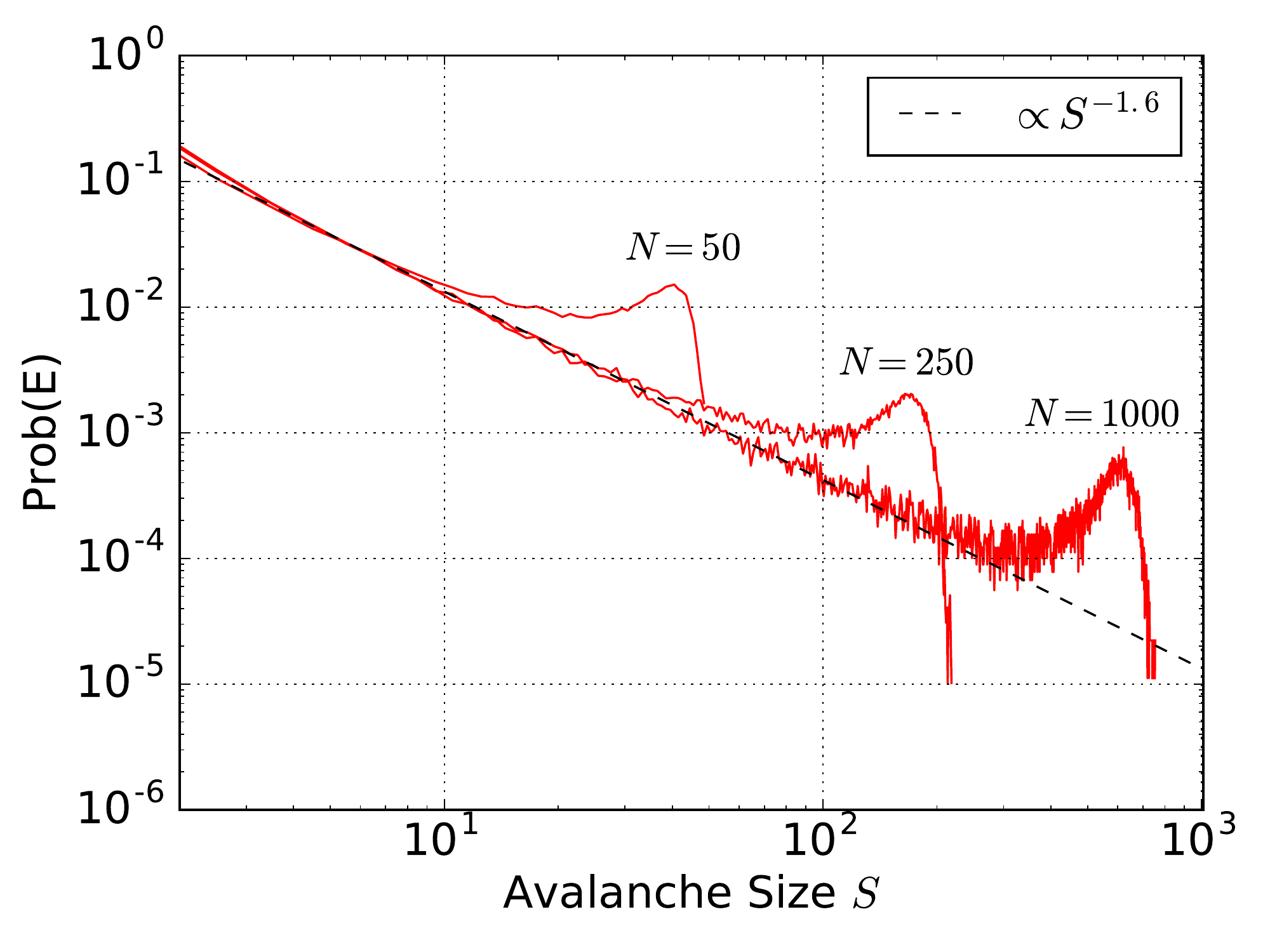}
	\caption{Scaling of the avalanche size distribution with increasing system size $N$. 
	Each distribution is obtained from 90000 avalanches. During one avalanche 
	each node can only contribute once to the avalanche size. Parameters: $\beta=10$, $W=1000$.}
	\label{FIG:Scaling of cut off}
\end{figure}

A further verification of criticality can be found in Figure \ref{FIG:Scaling of cut off} 
which shows the avalanche size distributions for different systems sizes $N$. 
With increasing system size the power-law-like regions of the distributions increase, 
showing that the cut-off is only a finite size effect. 
\section{Other versions of the model}
\label{Sec:Other Versions}
\label{SEC: Other versions}
The main goal of this work is to present a minimal adaptive network model which exhibits 
self-organized critical behavior. At the same time the model is supposed to be plausible, 
in the sense that only local information is used to approach the critical state. 
While we simulate this model on a von Neumann computer, a fully autonomous implementation is possible. 
To further demonstrate that our model represents a general mechanism and does not depend on particular
features of the implementation, also variants of the model were tested. 

\subsection{Inhibiting nodes}
We tested a variant which uses inhibiting nodes instead of inhibiting links. 
In this modified model, nodes are connected by un-weighted links. 
Before starting the evolution of the network a fraction of all nodes is chosen 
to be inhibitory. Here we typically chose 20--30~\% as it is often used as rough 
approximation for real neural systems \cite{Markram2004}. 
Simulations revealed that in the frame of the model the exact number is not of importance. 
 
If an inhibitory node is active, it contributes a signal $-1$ to the inputs of all nodes 
to which it is connected via outgoing links. Furthermore, the  second rule of the rewiring algorithm 
\begin{itemize}   
	\item   $A_i=1$: add a new incoming link $c_{ij}=-1$ from another randomly chosen node $j$       
\end{itemize}
is changed to
\begin{itemize}   
	\item   $A_i=1$: add a new incoming link from another randomly chosen inhibitory node $j$.       
\end{itemize}
Defining the rewiring rules in this way, the statistical properties of the inhibiting nodes 
do not differ from the activating nodes. Therefore, the dynamics of this modified version 
should be the same as the dynamics of the original model as, indeed, has been confirmed by 
simulations.

\subsection{Continuous link weights}
Choosing discrete link weights $c_{ij} \in \{ -1,0,1\} $ allows for a minimalistic description 
of the model and to formulate simple rules for the network evolution. 
However, in order to mimic the varying synaptic strengths of a real neural system, 
a version with continuous link weights has also been examined. 
We find that the following continuous rewiring rules lead to critical behavior, as well. 
In the same way as in the original model, after every $W$ time steps, one node $i$ is chosen at random. 
Depending on its average activity $A_i$ its linkage is changed as described in the following:
\begin{itemize} 
	\item $A_i=0$: randomly choose another node $j$. If $c_{ij} =0$ add a new incoming link $c_{ij}\in [0,\Delta]$. If $c_{ij} \ne 0$ multiply the link weight by a factor $(1+\delta \, \sign \, (c_{ij}) )$.
	\item   $A_i=1$:  randomly choose another node $j$. If $c_{ij} =0$ add a new incoming link $c_{ij}\in [-\Delta,0]$. If $c_{ij} \ne 0$ multiply the link weight by a factor $(1-\delta \, \sign \, (c_{ij}) )$.    
	\item   $A_i \not\in \{0,1\}$: randomly choose one incoming link of $i$. If $\left|c_{ij}\right|<1$ set $c_{ij}=0$, otherwise decrease the link weight by a factor $(1-\delta)$.
\end{itemize}
Hereby, the additional parameters $\delta$ and $\Delta$ should be chosen such that $\delta \ll 1$ 
to keep incremental changes small, and $\Delta >2$ for new links to have a dynamical effect 
in the face of the threshold update rule. Then the network robustly reaches a critical state. 

\section{Conclusion and outlook}
\label{Sec:Conclusion}
In this paper we studied a minimal neural network model that self-organizes towards a critical state, 
reproducing detailed features of criticality observed in real biological neural systems. 
The model involves only three parameters, the inverse temperature $\beta$ 
determining the amount of noise in the model, the averaging time $W$ defining the timescale 
on which the network evolution is performed, and the system size $N$. 
None of them is in need of fine tuning and they can be varied over a considerable range. 
We reformulated the earliest network SOC model \cite{Bornholdt2000} with Boolean variables
and modified the evolution rule accordingly, resulting in a mechanism which drives the autonomous
network model towards criticality whenever spontaneous activity is present. 
Note that this mechanism is based on the temporally averaged activity of single nodes only. 
Thus, it is shown that neither information about the global state of the network, 
nor information about neighboring nodes is necessary for obtaining neural networks 
showing self-organized critical behavior. 
 
In contrast to classical systems exhibiting self-organized criticality, as e.g.\ the sandpile model 
\cite{bak1988self}, the model shows critical features over a broad range of noise. 
Indeed, it even utilizes noise in order to sustain activity permanently. The source of robustness 
of this class of SOC models is the fact that the criticality of the system is stored in separate variables, 
namely the links between the nodes, rather than in the dynamical variables, the node states, themselves. 
Classical SOC models are more vulnerable against noise, as can be seen, for example, in the forest fire
model, where criticality emerges as a fractal distribution of tree states, that is easily disturbed. 
In our self-organized critical adaptive network model, in contrast, noise can be varied over a broad range. 

We have demonstrated that the rewiring mechanism also works in different versions of the model 
where, for example, inhibiting nodes instead of inhibiting links are implemented or continuous link 
weights are used. 
This illustrates that the observed self-organized critical characteristics arise as stable phenomena 
independent of even major features of the system, only depending on the structure of the rewiring algorithm.
This independence and robustness gives strong support to the hypothesis that also real biological neural systems 
could take advantage of this simple and robust way to self-tune close to a phase transition in order to 
stay away from, both, frozen as well as chaotic dynamical regimes.  

Further work on minimal neural network models showing critical behavior could focus on how 
criticality influences learning, as it already has been touched e.g.\ in \cite{Arcangelis2010,Hernandez-Urbina2016}. 
Further studies on self-organized critical networks models could not only help us in our understanding of 
biological neural systems, but also motivate new ways to generate artificial neural networks through morphogenesis.

\section*{Acknowledgments}
We thank Gorm Gruner Jensen for discussions and comments on the manuscript.
\bibliographystyle{ieeetr} 

\end{document}